\documentclass[aps,prb,twocolumn,superscriptaddress,showpacs]{revtex4-1}
\pdfoutput=1

\usepackage[bookmarks=false]{hyperref}

\usepackage{graphicx}
\usepackage{dsfont}
\usepackage{amsmath}
\usepackage{amsfonts}
\usepackage{amssymb}
\usepackage{amsthm}
\usepackage{xcolor}
\usepackage{soul}
\usepackage[caption=false]{subfig}
\usepackage[subnum]{cases}
\newcommand{\bra}[1]{\ensuremath{\left\langle#1\right|}}
\newcommand{\ket}[1]{\ensuremath{\left|#1\right\rangle}}
\DeclareMathOperator{\tr}{tr}
\usepackage{cleveref}

\begin{document}

\title{Efficiency of free auxiliary models in describing interacting fermions:\\ from the Kohn-Sham model to the optimal entanglement model}

\author{Kristian Patrick}
\email{py11kp@leeds.ac.uk}
\affiliation{School of Physics and Astronomy, University of Leeds, Leeds, LS2 9JT, United Kingdom}
\author{Marcela Herrera}
\affiliation{Centro de Ci\^encias Naturais e Humanas, Universidade Federal do ABC, Avenida dos Estados 5001, 09210-580 Santo Andr\'e, Sao Paulo, Brazil}
\author{Jake Southall}
\affiliation{School of Physics and Astronomy, University of Leeds, Leeds, LS2 9JT, United Kingdom}
\author{Irene D'Amico}
\email{irene.damico@york.ac.uk}
\affiliation{Department of Physics, University of York, York, YO10 5DD, United Kingdom}
\author{Jiannis K. Pachos}
\affiliation{School of Physics and Astronomy, University of Leeds, Leeds, LS2 9JT, United Kingdom}


\begin{abstract}
Density functional theory maps an interacting Hamiltonian onto the Kohn-Sham Hamiltonian, an explicitly free model with identical local fermion densities.
Using the interaction distance, the minimum distance between the ground state of the interacting system and a generic free fermion state, we quantify the applicability and limitations of the exact Kohn-Sham model in capturing the various properties of the interacting system.
As a byproduct, this distance determines the optimal free state that reproduces the entanglement properties of the interacting system as faithfully as possible.
The parent Hamiltonian of the optimal free state identifies a system that can determine the expectation value of any observable with controlled accuracy.
This optimal entanglement model opens up the possibility of extending the systematic applicability of auxiliary free models into the non-perturbative, strongly-correlated regimes.
\end{abstract}


\maketitle

\section{Introduction}

Undoubtably, interactions give rise to a wide range of quantum phases of matter with intriguing and exotic properties, ranging from many-body localisation~\cite{Huse} to the fractional quantum Hall effect~\cite{Laughlin}.
Nevertheless, the theoretical investigation of interacting systems is often formidable due to their complexity~\cite{Verstraete,Vidal,Nightingale}.
A possible approach in studying interacting systems is to approximate them by free models that offer a simpler and intuitive description.
To this aim,  physicists, chemists and material scientists alike often use Density Functional Theory (DFT)~\cite{HK,Gunnarsson,Jones,Capelle1}.

In its basic formulation, DFT uniquely maps a many-body system to an auxiliary non-interacting one, known as the Kohn-Sham (KS) model \cite{KS}, which has the same ground state fermion density as the interacting system.
In principle DFT ensures that any physical observable could be written as a functional of such ground-state density; in practice, with few exceptions, the forms of these functionals are unknown, and so it is often the case that the KS model itself is directly used as a non-interacting approximation to the many-body system, and the properties of the latter are then estimated by using the KS wave-functions in lieu of the many-body ones. Here, we consider the KS model in this acceptation.
 In this sense, and even with its known limitations in the strong correlation limit, the KS model has been used to estimate many-body properties, such as band-structure calculations~\cite{Capelle1,Capelle}, quantum work~\cite{Herrera} and entanglement~\cite{Coe:2008}.
As an attempt to further understand the entanglement properties of the KS model, current research focuses on certain models and the ability of KS to reproduce specific entanglement measures~\cite{Coe:2008}.
Alternatively, entanglement is used to enhance current DFT methods in order to find accurate ground states for very large system sizes~\cite{Wagner}.
However, it is not known how optimal the KS model is within the set of all possible free fermion theories.

To quantify the applicability of the KS model we employ the concept of interaction distance, $D_{\cal F}$~\cite{Turner}.
This distance measures how far the ground state of a given system is from the manifold of all free fermion states in terms of their quantum correlations across a geometric bipartition of the state~\cite{Turner,Kon,Jiannis}.
This should be contrasted to other approaches, such as $1-$particle Reduced Density Matrix~\cite{Lowdin}, where a restricted free fermionic manifold is implicitly assumed.
The interaction distance also identifies the optimal free state, a state with bipartite entanglement properties as close as possible to the interacting system.

We demonstrate that, in the perturbative (weak-interaction) regime, if $D_{\cal F}\approx 0$ then the KS ground state is close to the optimal free state, where the optimal free state has an error in determining {\it any} observable of the interacting system bounded by $D_{\cal F}$.
We also show that, away from the perturbative regime the reliability of the exact KS model as an approximation to the interacting model is limited to finding only the (exact) local densities, whereas large errors are associated to other quantities, such as the entanglement entropy, even if $D_{\cal F}\approx 0$.
Such a situation may appear e.g.~where interactions ``freeze" some fermionic degrees of freedom used to build the KS model.
To describe the interacting model as faithfully as possible in all coupling regimes we introduce the ``optimal entanglement model", with a Hamiltonian that has the optimal free state as its ground state.
We demonstrate that the optimal entanglement model reproduces all the ground state properties of the interacting system with an error bounded by $D_{\cal F}$, even in the strong-correlation regime.
This characteristic of the optimal entanglement model is not shared by the KS or other free approximations, e.g. Hartree-Fock, where restrictions over the form of the free fermion Hamiltonian, and/or the focus on optimizing quantities such as the particle density or the energy, limits their ability to accurately capture the entanglement resulting from strong interactions.

To exemplify this approach we employ the Fermi-Hubbard model restricted to $L=2$ sites -- the Hubbard dimer.
Due to its small system size, it is possible to analytically obtain the ground state of the interacting model, and thus determine the exact KS model and the optimal entanglement model.
As a result we can obtain a closed form for the interaction distance, $D_{\cal F}$, of the Hubbard dimer as a function of the interaction coupling $U$, and identify the regimes where the KS and the optimal entanglement models are good approximations to the interacting system and where their application is limited.
Our analytical treatment demonstrates that for the Hubbard dimer the interaction distance behaves like $D_{\cal F}\propto U^{-3}+{\cal O}(U^{-5})$ for large $U$. Hence, in the strongly correlated regime the ground state of the Hubbard dimer admits a free fermion description within an error that goes to zero as $U^{-3}$.
In the following section we introduce DFT and the interaction distance.

\section{Density Functional Theory and the Kohn-Sham Model}

Let us consider a Hamiltonian
\begin{align}
\hat{H} = \hat{K}+\hat{V}+\hat{W}
\end{align}
on a lattice, built from a kinetic energy operator $\hat{K}$, a local potential operator $\hat{V}=\sum_j^L v_j\hat{n}_j$, where $\hat{n}_j$ is the site-occupation operator, and a particle-particle interaction operator $\hat{W}$.
At the core of lattice-DFT are the one-to-one correspondences between the ground state wave function $|\psi\rangle$, the corresponding ground state density $\langle \hat{n}_j\rangle$ for ${j=1,\dots ,L}$, and the local potential~\cite{Coe} of an $L$-site many-body system~\cite{Capelle}.
With the ansatz of $n$ and $v$-representability, these correspondences imply that there exists a unique non-interacting model, the so called Kohn-Sham model~\cite{KS}, which is subject to the same kinetic operator and having the same ground-state density as the original $N$-particle interacting system.
Through this model, the density, and then in principle all other physical many-body properties~\cite{Capelle1}, can be obtained.
The KS Hamiltonian is given by
\begin{align}
\hat{H}_\text{KS}=\hat{K}+\hat{V}_\text{KS},
\end{align}
 where the potential $\hat{V}_\text{KS}$ is a combination of the original one-body potential, $\hat{V}$, the Hartree potential, representing the classical electrostatic interaction, and the exchange-correlation potential.
The latter contains contributions from the many-body interactions of the original system.
Apart from relatively simple systems, determining the KS model requires approximations \cite{Capelle1}.
Nevertheless, it is a significantly simpler task than solving the interacting system.

\section{The Optimal Free State and Optimal Entanglement Model}

\subsection{The Interaction Distance}
Let us now consider the entanglement properties of an interacting system.
For a given bipartition into a region $A$ and the complement $B$ of its ground state $\ket{\psi}$ the reduced density matrix is $\rho^\text{int}=\tr_B \ket{\psi}\!\bra{\psi}$, that has eigenvales $\lbrace \rho^\text{int}_k\rbrace$ related to the entanglement spectrum by $E^\text{int}_k=-\ln\rho^\text{int}_k$.
The entanglement entropy is defined as $S(\rho^\text{int})=-\tr \rho^\text{int}\ln\rho^\text{int}=-\sum^M_{k=1} \rho^\text{int}_k \ln \rho^\text{int}_k$, that is maximal $S_\text{max}=\ln M$ when the full set of $M$ entangled modes are equally weighted.

The interaction distance~\cite{Turner} of $\rho^\text{int}$ is defined as
\begin{equation}\label{eq:df}
D_{\mathcal{F}}(\rho^\text{int}) = \min\limits_{\rho^\text{free}\in\mathcal{F}}D_{\tr}\left(\rho^\text{int},\rho^\text{free}\right),
\end{equation}
where $D_{\tr}(\rho,\sigma)=\frac{1}{2}\tr\left|\rho - \sigma\right|$ is the trace distance metric between the reduced density matrices $\rho$ and $\sigma$, and the minimisation is over the whole set $\mathcal{F}$ of possible Gaussian states $\rho^\text{free}$.
This distance measures how distinguishable a fermionic state is from being free in terms of its ground state correlations across a bipartition.
It is often amenable to analytical calculations~\cite{Jiannis} and it can be numerically evaluated efficiently for 1D interacting systems with DMRG methods~\cite{Turner,Kon}.
We denote by $\rho^\text{opt}$ the optimal free density matrix that minimises the trace distance $D_{\tr}(\rho^\text{int},\rho^\text{free})$, and thus reproduces the entanglement properties of $\rho^\text{int}$ as faithfully as possible.
Its parent Hamiltonian is generally unknown and may offer complementary information with respect to the KS Hamiltonian that optimises over the local fermion densities.


\subsection{Bounding Observables with $D_\mathcal{F}$}
As the trace distance is the maximum distance over all positive operator valued measures~\cite{Nielsen}, we expect the state that minimises it over all free states to best approximate not only its bipartite entanglement, but also all other observable quantities.
Consider the expectation value of an observable $\mathcal{O}$ for two density matrices $\rho$ and $\sigma$ given by $\left\langle \mathcal{O}\right\rangle_\rho = \tr\left[ \mathcal{O}\rho\right]$ and $\left\langle \mathcal{O}\right\rangle_\sigma = \tr\left[ \mathcal{O}\sigma\right]$, respectively.
To compare these expectation values we define their difference by the metric
\begin{align}
d_\mathcal{O}(\rho,\sigma) = \left| \left\langle \mathcal{O}\right\rangle_\rho - \left\langle \mathcal{O}\right\rangle_\sigma\right|,
\end{align}
which reduces to $d_\mathcal{O} = \left| \tr\left[ \mathcal{O}(\rho-\sigma)\right]\right|$.
Let us express $\rho-\sigma$ in its diagonal basis, $\rho-\sigma= \sum_k \phi_k \ket{\phi_k}\!\!\bra{\phi_k}$, where $\phi_k$ are the eigenvalues of $\rho-\sigma$.
Then, via direct substitution into $d_\mathcal{O}$, we find that
\begin{align}
d_\mathcal{O}&=\left|\tr\left[\mathcal{O}\sum_k \phi_k \ket{\phi_k}\!\!\bra{\phi_k}\right]\right|\\
&=\left|\sum_k\bra{\phi_k}\mathcal{O}\ket{\phi_k} \phi_k\right|\\
&\leq \left|\max\limits_k \bra{\phi_k}\mathcal{O}\ket{\phi_k} \sum_k \phi_k\right| = \left|\mathcal{O}_\text{max}\right|\left|\sum_k \phi_k\right|,
\end{align}
where $\mathcal{O}_\text{max} $ is the largest eigenvalue of the operator $\mathcal{O}$ in absolute value.
It then follows that
\begin{align}
d_\mathcal{O} &\leq \left|\mathcal{O}_\text{max}\right| \sum_k \left|\phi_k\right|= \left|\mathcal{O}_\text{max}\right| \tr \left|\rho-\sigma\right|,
\end{align}
where the final equality explicitly contains the definition of the interaction distance when $\sigma = \rho^\mathrm{opt}$.
Therefore, when $\rho = \rho^\mathrm{int}$ and $\sigma = \rho^\mathrm{opt}$ the difference in expectation values are bounded by the interaction distance, i.e.
\begin{align}\label{eqn:ineq}
\left| \left\langle \mathcal{O} \right\rangle_{\rho^\text{int}} - \left\langle \mathcal{O}\right\rangle_{\rho^\text{opt}} \right|\leq C_\mathcal{O} D_\mathcal{F},
\end{align}
with $C_\mathcal{O} = \frac{1}{2} \left|\mathcal{O}_\text{max}\right|$ that depends only on the operator $\mathcal{O}$. As a result, the expectation value of any observable $\mathcal{O}$ with respect to the ground state $\rho^\text{int}$ of the interacting system can be reproduced by the optimal free state $\rho^\text{opt}$ with an accuracy that is controlled by $D_\mathcal{F}$.
In contrast to Eq.~\eqref{eqn:ineq}, other methods aim to optimally determine a single observable at the expense of introducing unbounded error on the rest of the complementary observables \cite{Bach}.
This is the case for DFT, as explicitly shown for in Fig.~\ref{fig:distanceplots}.

\subsection{Bounding Density with $D_\mathcal{F}$ }

We would like now to compare the applicability of the KS ground state and the optimal free state.
Let us apply inequality~\eqref{eqn:ineq} to the local density of fermions, ${\cal O}=\hat{n}_{j}$.
For a state with reduced density matrix $\rho$ at site $j$ we define $\hat{n}_{j,\rho}=\tr(\rho\hat{n}_j)$.
The `natural' metric~\cite{Damico}, between $\rho^\text{int}$ and $\rho^\text{opt}$, on the metric space of local densities over all sites is given by
\begin{align}
D_n(\rho^\text{int},\rho^\text{opt}) =  \sum_j \left|\hat{n}_{j,\rho^\text{int}}-\hat{n}_{j,\rho^\text{opt}}\right|.
\end{align}
To arrive at this definition from Eq.~\eqref{eqn:ineq}, we must sum over all sites.
Then, Eq.~\eqref{eqn:ineq} becomes
\begin{align}\label{eq:appmetricbound}
\sum_j\left| \left\langle \hat{n}_j \right\rangle_{\rho^\text{int}} - \left\langle \hat{n}_j\right\rangle_{\rho^\text{opt}} \right|\leq \sum_j C_{\hat{n}_j} D_\mathcal{F}.
\end{align}
The left hand side of this equality is the definition of the natural metric and the right hand side consists of a constant $C = \sum_j 	C_{\hat{n}_j}$ multiplied by the interaction distance.
The bound reduces to
\begin{align}
D_n(\rho^\text{int},\rho^\text{opt})\leq C D_\mathcal{F}.
\end{align}
Due to the key property of the Kohn-Sham model, that $ \left\langle \hat{n}_j \right\rangle_{\rho^\text{int}} =  \left\langle \hat{n}_j \right\rangle_{\rho^\text{KS}}$, the bound may be cast in terms of the optimal and Kohn-Sham ground states
\begin{align}\label{eq:bound}
D_n(\rho^\text{KS},\rho^\text{opt})\leq C D_\mathcal{F}.
\end{align}
Hence, the interaction distance bounds the density distance between the KS and optimal free state.
This bound implies that for $D_{\cal F}\approx 0$, e.g. in the perturbative regime,
the optimal free state has fermion densities that are very close to the densities of the KS ground state.

\subsection{Trace Distance Bounding in Perturbative Limit}

We now investigate when the KS model reproduces also the entanglement properties of the optimal entanglement model.
Assume that the density matrices are a continuous functional of the fermion densities, e.g. when the system is in the perturbative regime with no phase transitions caused by the interactions. We can write $n_F = n +\delta n$, with $n_F$ the ground state density of the optimal free state, $n$ the ground state density of the interacting/KS model, and $\delta n$ a small linear response.

First consider the limit $\delta n\to0$.
In this limit $D_\mathrm{tr}\left(\rho^\mathrm{KS},\rho^\mathrm{opt}\right)\to0$ and $D_\mathcal{F}\to0$, so that the inequality above is satisfied by the equality $0=C\cdot 0$.
Next, consider the linear response to be small and non-zero.
From Eq.~\eqref{eq:bound}, it can be seen that the density metric is bound by the interaction distance.
When DFT Hohenberg-Kohn-type theorems apply, any property of a pure state interacting system described by a Hamiltonian $\hat{H} = \hat{K} +\hat{W} + \hat{V}$, can be written as a functional of the system ground state density.
So, in particular, the (non-diagonal) density matrix elements can also be written as a functional of the ground state density, and thus as a functional of $n$ and $\delta n$.
For small $\delta n$, we can approximate $D_\mathrm{tr}\left(\rho^\mathrm{opt},\rho^\mathrm{KS}\right)$ through a Taylor expansion around $\delta n=0$. Up to $\mathcal{O}(\delta n^2)$, the trace distance becomes
\begin{align}\label{eq:traceapprox}
D_\mathrm{tr}\left(\rho^\mathrm{KS},\rho^\mathrm{opt}\right) [\delta n,n] &\approx\left.\frac{\delta^2 D_\mathrm{tr}}{\delta n^2}\right|_{\delta n=0}(\delta n)^2>0,
\end{align}
which holds due to $\delta n =0$ being a minimum (and the trace distance being a metric).
Similarly, we can approximate the density metric about the minimum:
 \begin{align}
D_\mathrm{n}\left(\rho^\mathrm{KS},\rho^\mathrm{opt}\right)&= D_\mathrm{n}\left(\rho^\mathrm{KS},\rho^\mathrm{opt}\right)[\delta n,n] \\
&\approx\left. \frac{\delta^2 D_\mathrm{n}}{\delta n^2}\right|_{\delta n=0}(\delta n)^2>0\label{eq:densityapprox}
\end{align}
Using Eqs.~\eqref{eq:traceapprox} and \eqref{eq:densityapprox}, and up to higher orders than $(\delta n)^2$ in $\delta n$, we can then write
\begin{align}
D_\mathrm{tr}&\left(\rho^\mathrm{KS},\rho^\mathrm{opt}\right) [\delta n,n] \approx f(n)\cdot D_\mathrm{n}\left(\rho^\mathrm{KS},\rho^\mathrm{opt}\right),
\end{align}
where $f(n)=\left.\frac{\delta^2D_\mathrm{tr}}{\delta n^2}\right|_{\delta n=0}\left(\left.\frac{\delta^2D_\mathrm{n}}{\delta n^2}\right|_{\delta n=0} \right)^{-1}$ is a functional of $n$, but for a given $n$ it will be a number greater than zero.
Using Eq.~\eqref{eq:bound} we then obtain
\begin{align}
D_\mathrm{tr}\left(\rho^\mathrm{KS},\rho^\mathrm{opt}\right) \leq f(n) \cdot C D_\mathcal{F}.
\end{align}

Therefore, when the interaction distance is small then $\rho^\text{opt}$ and $\rho^\text{KS}$ are nearly overlapping and exhibit very similar entanglement properties.
Hence, in the perturbative regime for $D_{\cal F}\approx 0$ the KS model offers a way to constructively obtain the optimal free state.

\subsection{Triangle Inequality}
\begin{figure}[t]
{\includegraphics[width=0.9\linewidth]{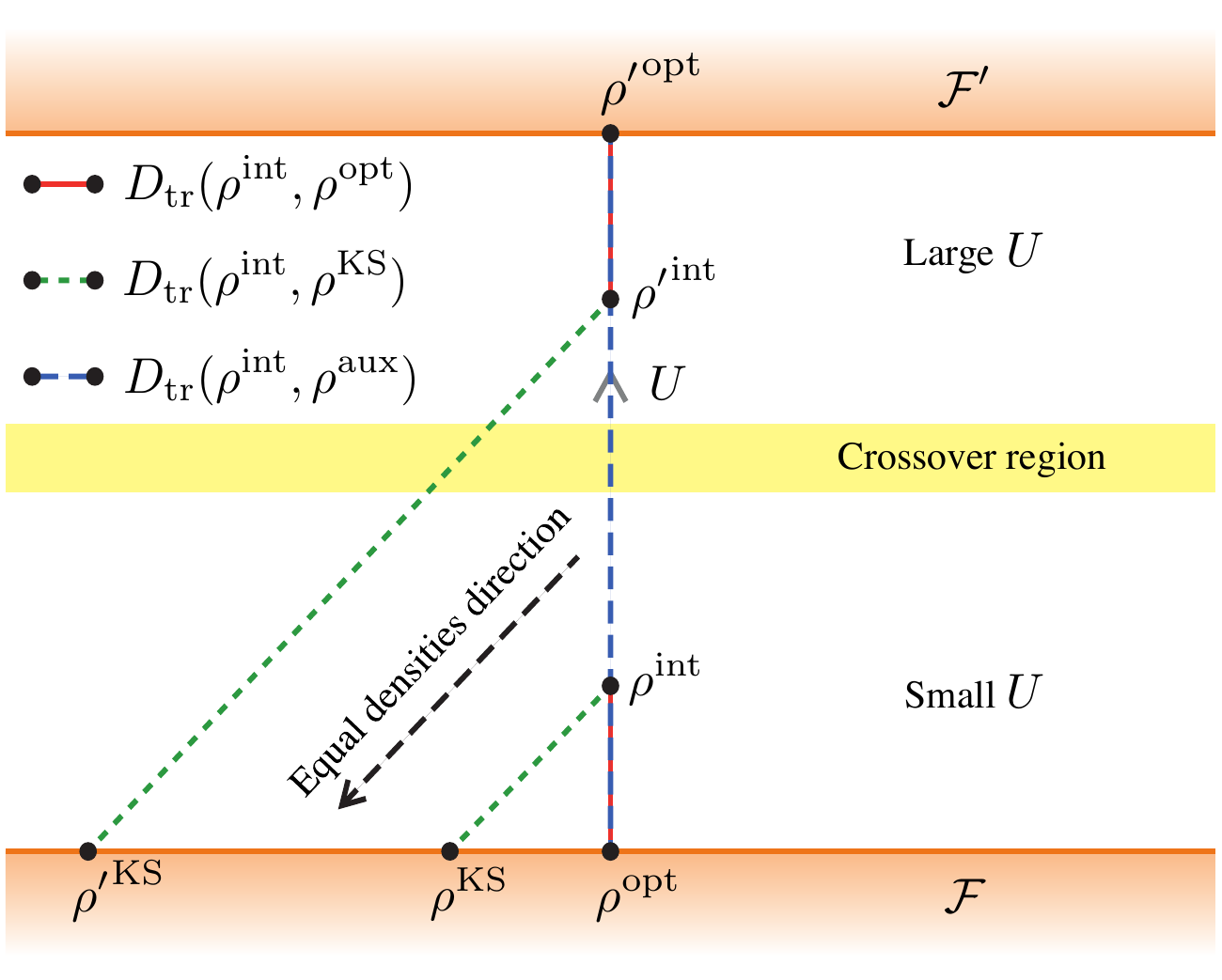}}
\caption{
A schematic illustration of the distances between interacting, $\rho^\text{int}$, optimal, $\rho^\text{opt}$, and Kohn-Sham, $\rho^\text{KS}$, reduced density matrices.
Two free manifolds of Gaussian states, ${\cal F}$ and ${\cal F}'$, are depicted with possibly different number of degrees of freedom.
For small interaction coupling $U$, $\rho^\text{int}$ is close to ${\cal F}$, while as $U$ increases ${\rho'}^\text{int}$ can be close to another manifold ${\cal F}'$.
The direction of equal local fermion densities identifies the KS model on ${\cal F}$.
In the perturbative regime (small $U$), when $D_{\mathcal{F}} \approx 0$, then $D_\text{tr}(\rho^\text{int},\rho^\text{KS})$ and $D_\text{tr}(\rho^\text{KS},\rho^\text{opt})$ are also small, as dictated by (\ref{eqn:bound1}).
When $U$ is large then the state ${\rho'}^\text{int}$ can effectively admit an optimal free description, ${\rho'}^\text{opt}$, with a different number of fermions than the one from the perturbative regime. This change makes the Kohn-Sham model, ${\rho'}^\text{KS}$, unsuitable for reproducing the entanglement properties of ${\rho'}^\text{int}$.
\label{fig:geometry1}
}
\end{figure}

We now investigate the bipartite entanglement of the model. 
We employ the triangular inequality of the trace distance metric between the interacting, $\rho^\text{int}$, the optimal free, $\rho^\text{opt}$, and the KS, $\rho^\text{KS}$, reduced density matrices, as shown in Fig.~\ref{fig:geometry1}.
As, in the perturbative regime, the interaction distance provides an upper bound for $D_{\tr}(\rho^\text{KS},\rho^\text{opt})$ we have $D_{\tr}(\rho^\text{int},\rho^\text{KS}) \leq (1+c) D_{\mathcal{F}}$.
Moreover, due to the optimality of $\rho^\text{opt}$ we have that $D_{\cal F}$ also lower bounds $D_{\tr}(\rho^\text{int},\rho^\text{KS})$, thus giving
\begin{align}
D_{\cal F} \leq D_{\tr}(\rho^\text{int},\rho^\text{KS}) \leq (1+c) D_{\mathcal{F}}.
\label{eqn:bound1}
\end{align}
Hence, in the perturbative regime when $D_{\cal F}\approx 0$ the KS model faithfully reproduces all the properties of the interacting system, while a non-zero $D_{\cal F}$ bounds the errors in determining the entanglement properties of the interacting model.
Away from the perturbative regime it is possible that the upper bound in~\eqref{eqn:bound1} fails, by having $\rho^\text{KS}$ far from $\rho^\text{int}$ even if $D_{\cal F}\approx 0$, as shown in Fig.~\ref{fig:geometry1}.
Nevertheless, $\rho^\text{opt}$ would still provide a faithful description of $\rho^\text{int}$.


The parent Hamiltonian of the optimal free state can be used to define a suitable auxiliary free model that identifies the effective degrees of freedom of the interacting model for all coupling regimes.
When $D_{\cal F}\approx 0$ such an auxiliary model not only faithfully reproduces the entanglement properties of the interacting model but, due to Eq.~\eqref{eqn:ineq}, it can also estimate all of its observables, such as the local fermion densities.
This `optimal entanglement' model generalises the KS model that can fail to reproduce the entanglement properties even if $D_{\cal F}\approx 0$.
In fact, strong interactions may not only change the effective local fermion potential, $\hat{V}$, but also the kinetic term, $\hat{K}$.
To build this auxiliary model one first needs to identify the effective fermionic degrees of freedom that correspond to the quantum correlations of the model.
If $D_{\cal F}\approx 0$ for strong interactions then the number of fermionic degrees of freedom of the emerging free theory can be either the same or smaller than the initial theory without the interaction term: interactions could freeze some of the initial fermionic degrees of freedom but they cannot increase their number.
To exemplify this procedure we apply it next to the Fermi-Hubbard model at half-filling, restricted to $L=2$ sites.

\section{The Fermi-Hubbard Model}

The 1D Hubbard model \cite{Hubbard} has successfully reproduced a number of physical phenomena, including interaction-driven quantum phase transitions~\cite{Essler}.
In some limits it has exact solutions \cite{Lieb,Carmelo} and has been studied via many numerical techniques including DFT~\cite{Capelle}.
It comprises spin-$\frac{1}{2}$ fermions with a creation (annihilation) operator $c^{\dagger}_{j,\sigma}$ ($c_{j,\sigma}$) at site $j$ and spin $\sigma\in\{\uparrow,\downarrow\}$, with Hamiltonian
\begin{align}
\label{eq:hamiltonian}
\hat{H} = \sum_{\substack{j,\sigma}}
\left[-J \left(c^{\dagger}_{j,\sigma}c^{}_{j+1,\sigma} + \text{h.c.}\right) + \nu_j \hat{n}_{j,\sigma} \right]+ U\sum_j \hat{n}_{j,\uparrow}\hat{n}_{j,\downarrow}
\end{align}
where $\hat{n}_{j,\sigma} = c^{\dagger}_{j,\sigma}c^{}_{j,\sigma}$ is the number operator, $J$ is the tunnelling strength, $U$ is the on-site particle-particle interaction strength, and  $\nu_j$ is the site-dependent potential.
At half-filling $N_\uparrow=N_\downarrow=L/2$, the model in the thermodynamic limit has two phases: for $U=0$ it is described by the Luttinger liquid phase, where local fermion densities are free to change, and for $U>0$ it is described by the Mott-insulator phase, where local densities are frozen by Coulomb repulsion~\cite{Essler}.
For finite system sizes and with anisotropic local potentials, the fluid phase extends into some small range of interaction strengths, leading to a `crossover region'.
Hence, it is an ideal system to demonstrate the applicability of the optimal entanglement model.

To study in detail the efficiency of the KS and the optimal entanglement models in representing the interacting ground state, we focus on the half-filled Hubbard dimer ($L=2$). This model enjoys analytical solutions for the ground state $\rho^\text{int}$ and its energy $E$~\cite{Carrascal}.
For this system size the KS model can be numerically determined exactly.
We can also derive exact solutions for the optimal free state of a four dimensional $\rho^\text{int}$, as is the case of the Hubbard dimer when restricted to zero total spin subspace (see Appendix).
As a result the interaction distance of the ground state in the strongly correlated regime is given by
\begin{equation}
D_\mathcal{F} = \frac{2J^2}{N} \left| \frac{(U-\Delta\nu-E)^2-(U+\Delta\nu-E)^2}{(U-\Delta\nu-E)^2}\right|
\label{eqn:Dff}
\end{equation}
where $N=4J^2+2(U+\Delta\nu-E)^2+4J^2(U+\Delta\nu-E)^2/(U-\Delta\nu-E)^2$, $\Delta\nu=\nu_1-\nu_2$ the asymmetry of local potentials and $E$ is its energy eigenvalue.
The large $U$ limit expansion of (\ref{eqn:Dff}) for constant $\Delta\nu$ and $J$ is given by
\begin{align}
D_\mathcal{F} = \frac{4J^2\Delta\nu}{U^3}+\mathcal{O}(U^{-5}),
\label{eqn:error}
\end{align}
demonstrating that the interaction distance rapidly approaches zero as $U$ increases, while it becomes truly free at $U= \infty$. In conclusion, as deduced from (\ref{eqn:ineq}), any observable has a ground state expectation value that can be approximated by the optimal free fermion state with an error given by (\ref{eqn:error}).
Moreover, we can analytically determine the optimal entanglement model, that reproduces exactly the entanglement spectrum of the optimal free state in the insulating phase, where we expect the exact KS model to become a bad approximation. We analyse this in detail below.

\subsection{An Optimal Entanglement Model for the Hubbard Dimer at Half-filling}
When the Hubbard model is restricted to two sites at half-filling the Hilbert space is spanned by the basis $\{\ket{\uparrow\downarrow,0},\ket{\uparrow,\downarrow},\ket{\downarrow,\uparrow},\ket{0,\uparrow\downarrow}\}$, where the basis state $\ket{x,y}=\ket{x}\otimes\ket{y}$ corresponds to $x$ fermions at the first site and $y$ fermions at the second with the indicated spin $\uparrow$ or $\downarrow$.
Eigenstates of this Hamiltonian have both occupation and spin degrees of freedom that can be varied by tuning the tunnelling and repulsive interaction strength.
By observation of the optimal free state entanglement spectrum in the insulating phase, found using the exact solutions from the appendix, it can be seen that there exists a double degeneracy.

In order to reproduce this optimal free entanglement spectrum, we construct an auxiliary model with two spinless non-interacting fermions hopping on separate two site chains.
Then, by appropriately tuning a chemical potential, $\mu$, on a single site to imitate the affect of interactions, it is possible to match exactly the double degeneracy of the optimal free entanglement spectrum.
This is akin to DFT where interaction effects are tuned through a potential to find accurate local densities; however, here we tune a potential to produce the optimal free entanglement spectrum that results in a controlled error over all observable quantities, as shown in Eq.~\eqref{eqn:ineq}.
Such a spectrum can be reproduced by the following Hamiltonian:
\begin{align}\label{eq:dimeraux}
\hat{H}_\mathrm{aux} = -J\left( c^\dagger_{1}c_{3} +c^\dagger_{3}c_{1}\right) -J \left( c^\dagger_{2}c_{4} +c^\dagger_{4}c_{2}\right) - \frac{\mu}{2} c^\dagger_{1}c_{1}
\end{align}
where $\mu = 2\left[\sqrt{\frac{\rho^\mathrm{opt}_1}{\rho^\mathrm{opt}_2}} - \sqrt{\frac{\rho^\mathrm{opt}_2}{\rho^\mathrm{opt}_1}}\right]$ and $(\rho^\mathrm{opt}_1,\rho^\mathrm{opt}_2)$ are the two distinct optimal entanglement levels.
The partition that returns the desired spectrum separates sites $1,2$ into subsystem $A$ and sites $3,4$ into subsystem $B$.
We note that~\eqref{eq:hamiltonian} and~\eqref{eq:dimeraux} have, at half filling, the same number of degrees of freedom: in particular this means that the auxiliary system~\eqref{eq:dimeraux} can reproduce the overall behaviour of both spin and charge degrees of freedom of~\eqref{eq:hamiltonian}.

\begin{figure}[t]
\centering
\setbox1=\hbox{\includegraphics[height=2cm]{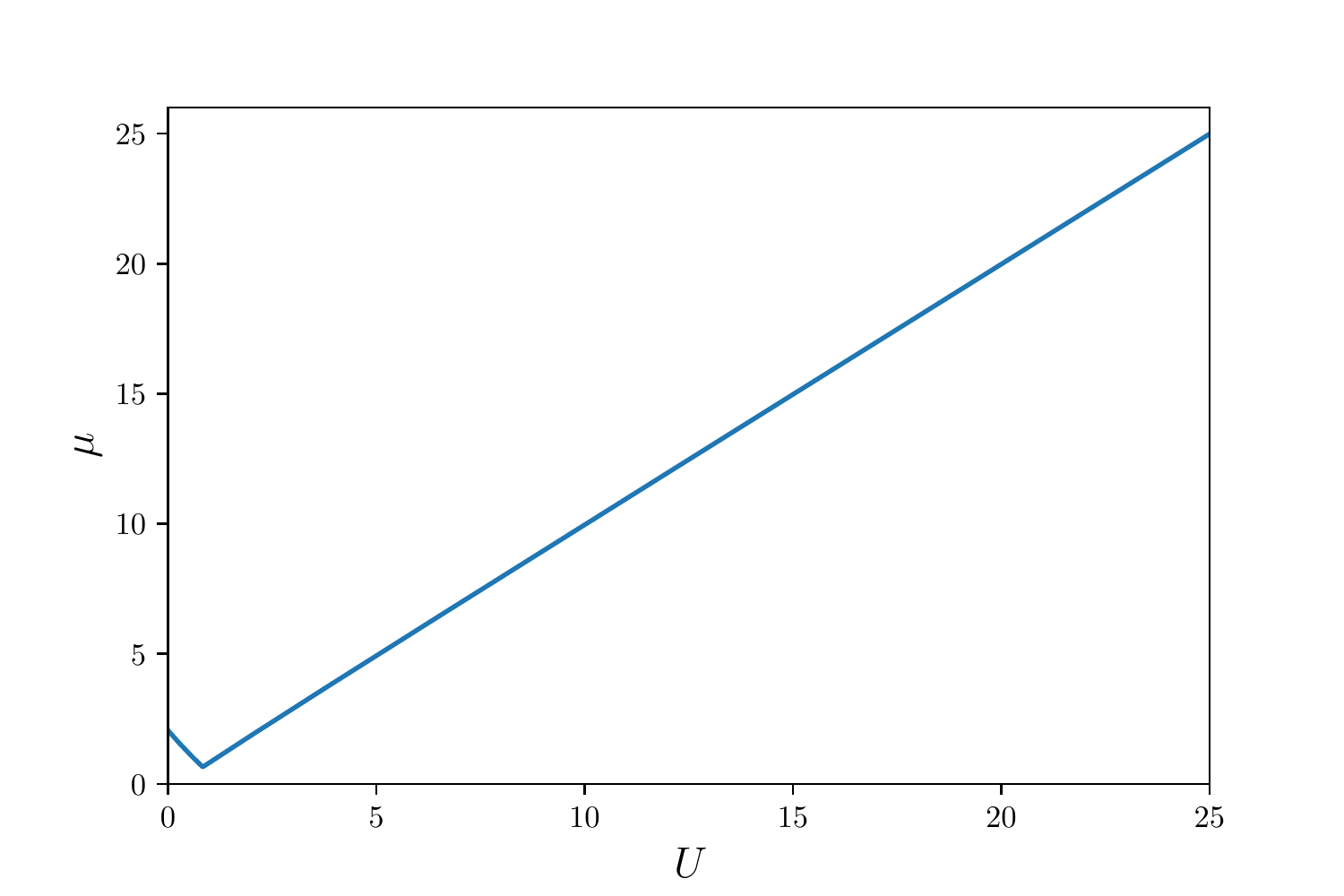}}
\includegraphics[width=0.45\paperwidth]{auxconstant}\llap{\makebox[\wd1][l]{\raisebox{3.75cm}{\hspace{-50mm}\includegraphics[height=1.5cm]{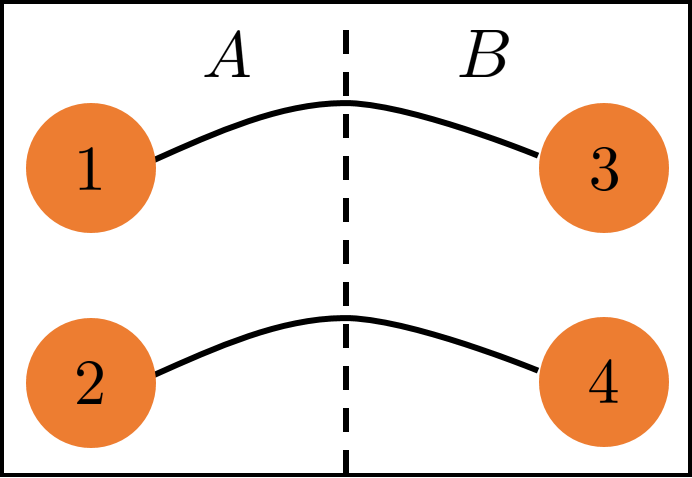}}}}
\caption{
Behaviour of the chemical potential $\mu$, for the $L=2$ optimal entanglement mode at half-filling, against interaction strength of the interacting model, $U$.
This chemical potential imitates the affect of interactions in the interacting model and its explicit form is a function of the optimal free state entanglement spectrum.
In the strong interaction regime we find $\mu\approx JU$ to a very good approximation.
Inset: A sketch of the auxiliary model described by Hamiltonian~\eqref{eq:dimeraux}.
The system is built from two non-interacting chains, each with a single spinless fermion.
The dashed line shows the partitioning of the system into subsystems $A$ and $B$.
}
\label{fig:auxconstantscaling}
\end{figure}

As the optimal free entanglement levels are functions of the couplings of the interacting model we have that $\mu=\mu(J,\nu_j,U)$.
Further, as shown in Fig.~\ref{fig:auxconstantscaling}, we observe a linear behaviour $\mu\approx JU$ within the insulating phase.
Armed with an optimal entanglement model, we now present numerical results demonstrating its applicability compared to the exact KS model.

\begin{figure}[t]
\centering
\includegraphics[width=0.4\paperwidth]{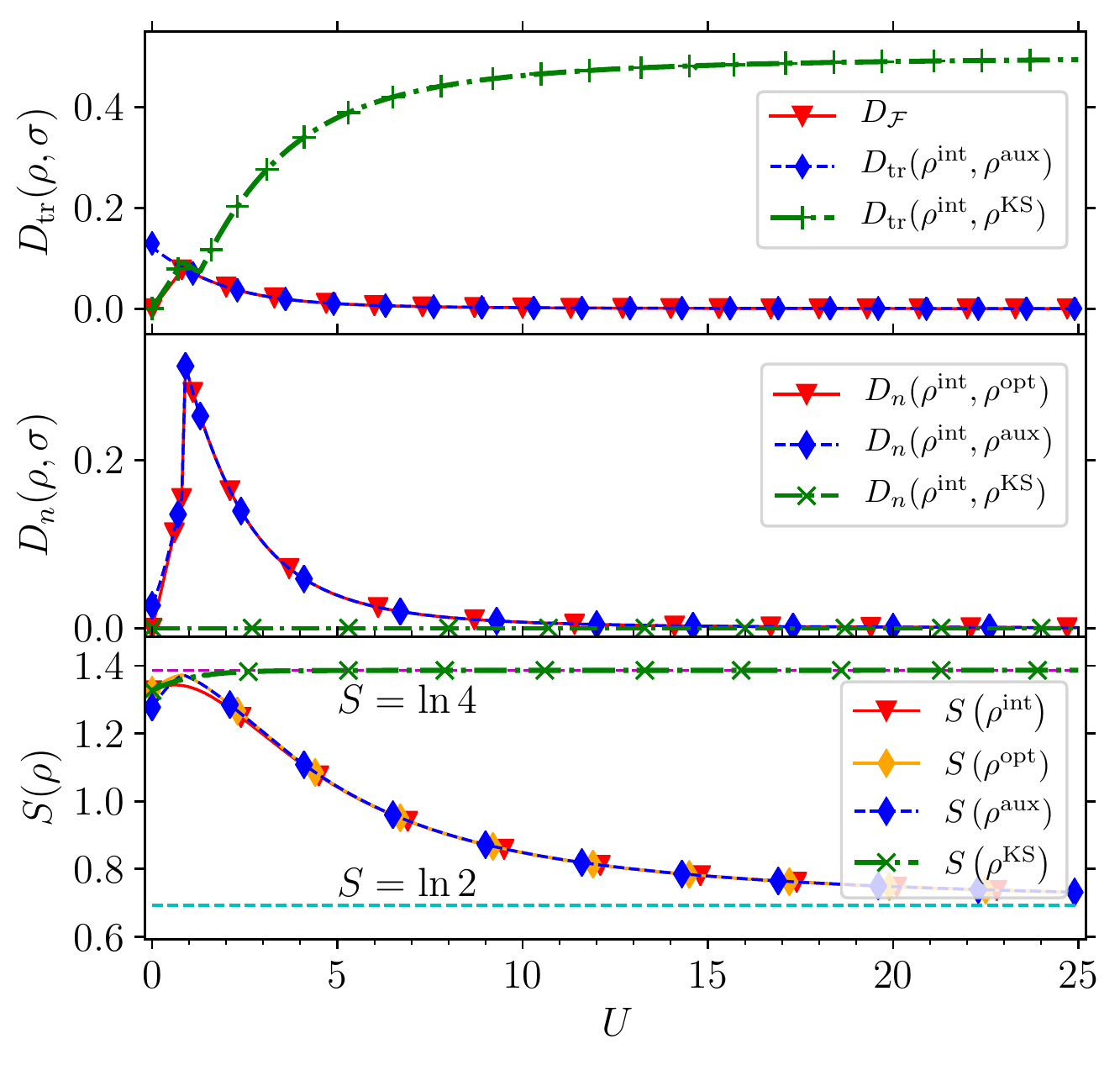}
\caption{
(Top) Trace distance, (Middle) natural metric, and (Bottom) entanglement entropy for the interacting, $\rho^{\mathrm{int}}$, optimal, $\rho^{\mathrm{opt}}$, KS, $\rho^{\mathrm{KS}}$, and auxiliary, $\rho^{\mathrm{aux}}$, reduced density matrices, as a function of the interaction coupling $U$, for $L=2$, $J=1$, total spin $S_z=0$, and $\nu_1 - \nu_2= 0.5$.
In the perturbative limit the KS is a good approximation to the optimal entanglement model which describes spin-$1/2$ free fermions.
In the large $U$ limit the KS model failsto reproduce entanglement, while both the optimal and auxiliary states that describe spinless free fermions, provide faithful representations of the local densities (Middle) and the entanglement entropy (Bottom) of the interacting system.
For large $U$ the entanglement entropy of the KS model tends to $S=\ln 4$
corresponding to the maximally entangled state $\ket{\psi}=(\ket{\uparrow\downarrow,0}+\ket{\uparrow,\downarrow}+\ket{\downarrow,\uparrow}+\ket{0,\uparrow\downarrow})/2$ while the interacting, optimal and auxiliary systems tend to $S\approx\ln 2$
that correspond to $\ket{\psi}=(\ket{\uparrow,\downarrow}+\ket{\downarrow,\uparrow})/\sqrt{2}$, signalling the freezing of double occupations due to interactions.
}
\label{fig:distanceplots}
\end{figure}

\subsection{Numerical Results}

The behaviour of the corresponding ground state reduced density matrices $\rho^\text{int}$, $\rho^\text{KS}$, $\rho^\text{opt}$, and $\rho^\text{aux}$, are given in Fig.~\ref{fig:distanceplots}.
Note that $D_{\cal F}$ is negligble for all values of $U$ away from the critical region $U\approx J$.
Surprisingly, the KS ground state closely approximates the optimal free state not only in the perturbative, but also in the intermediate coupling regime, $U\sim |J|,|\nu_j|$, up to the crossover.
Here, the trace distance between these states is small, so the KS is both exact in fermion density and also reproduces the ground state correlations of the optimal model accurately.
In the strong coupling regime, $U\gg |J|,|\nu_j|$, the KS model fails to reproduce the correlation properties of the optimal entanglement model. As it still describes correlations between spinful free fermions, it cannot accurately capture the entanglement of the interacting model.
This is in contrast to the optimal entanglement model that, in that regime, is described by spinless free fermions.
These degrees of freedom faithfully capture the quantum correlations of the interacting model, as shown in Fig.~\ref{fig:distanceplots} (Bottom).
Nevertheless, they only approximate its local densities, as shown in Fig.~\ref{fig:distanceplots} (Middle), with an error that is bounded by the value of $D_{\cal F}$, as dictated by Eq.~\eqref{eq:bound}. The local densities identify the change of the degrees of freedom from one optimal model to the other via the observed kink.

From the properties of the optimal free state we see that the effect of the strong interactions, in the $U\to\infty$ limit, is to freeze the local fermion populations to $n_j=1$ as an eigenvalue of the local density operator.
This can be witnessed by the behaviour of the entanglement entropy, $S$.
In Fig.~\ref{fig:distanceplots} we observe that the KS model saturates to the value $S=\ln 4$ due to both spin and population fluctuations.
In contrast, the interacting model has entanglement entropy that tends to $S=\ln 2$ as $U\to\infty$, due to only spin correlations.\\

The interaction distance is approaching zero for large $U$, signalling that the spin correlations can be faithfully reproduced by free fermions.
In this case, the optimal entanglement model with Hamiltonian~\eqref{eq:dimeraux} faithfully reproduces both the local densities as well as the correlation properties of the interacting system, as shown in Fig.~\ref{fig:distanceplots}.
Hence, unlike the KS model, it provides a faithful representation of the interacting theory.


To schematically present why the optimal entanglement model succeeds in faithfully representing {entanglement properties of} the interacting system for large $U$, while the KS model fails, we refer to the schematic in Fig.~\ref{fig:geometry1}. From the above analysis of the dimer model we observe that interactions have the effect of moving the optimal free state from describing free spinful fermions (manifold $\mathcal{F}$) towards the description of free spinless fermions (manifold $\mathcal{F}'$).
Due to the fixed form of the kinetic term of the KS Hamiltonian, its corresponding reduced density matrix will always live in $\mathcal{F}$.
By choosing the auxiliary model to optimise over entanglement, its degrees of freedom can change from ${\cal F}$ to ${\cal F}'$ that better describes the interacting system at large couplings $U$.
Thus, the optimal entanglement model is able to reproduce all the properties of the interacting system for all $U$, with an error that is bounded by $D_{\cal F}$.

\section{Conclusions}

With the help of the interaction distance, $D_{\cal F}$, we are able to identify the free model that approximates the interacting system by optimising over the corresponding entanglement properties.
We demonstrate that when the interaction distance is small then the optimal entanglement model reproduces all observables of the interacting system with accuracy bounded by $D_{\cal F}$.
As such, it provides an accurate modelling of the low energy behaviour of the system~\cite{Li,Turner}.
The exact KS model, on the other hand, finds local densities exactly for all strengths of interactions, but can dramatically fail to obtain entanglement features even when the interaction distance is small.
Motivated by these results we envisage that a method inspired by DFT, where the optimisation of the free model is performed with respect to entanglement properties rather than local densities, can faithfully approximate strongly interacting systems.

The idea of optimising DFT calculations over entanglement is not entirely novel.
The authors of~\cite{Wagner} show that, through a combination of DMRG and DFT, it is possible to obtain an accurate approximations to 3D physical systems through 1D calculations.
DMRG inherently optimises over entanglement and is advantageous here as it allows one to approach the continuum limit faster than through a direct study of a large strongly correlated physical system.
However, what we propose is unique as the optimal entanglement model reproduces the interacting state within the bound of the interaction distance.

To exemplify the diagnostic power of the interaction distance, we considered the Fermi-Hubbard dimer.
By studying its ground state entanglement spectrum we identified the small range of $U$'s up to the crossover region where the KS model approximates well the optimal entanglement model.
Beyond the crossover, and when in an insulating phase, the KS model entanglement spectrum diverges from the optimal entanglement model due to the fixed kinetic operator, as evidenced by its entanglement entropy.
The optimal entanglement model is defined to have the optimal free state as its ground state.
From it, it is possible to obtain all properties within the bound of the interaction distance for any choice of interaction strength.
In future work, we aim to generalise DFT with entanglement for larger system sizes of the Fermi-Hubbard model.

\begin{acknowledgements}
We would like to thank Pasquale Calabrese, Konstantinos Meichanetzidis, Zlatko Papic, and Christopher J. Turner for inspiring conversations.
Fig.~$1$ was designed by Jack White.
KP, JS, and JKP acknowledge support by the EPSRC grant EP/I038683/1.
MH acknowledges support from FAPESP (grant no.2014/02778-1)
IDA and MH acknowledge support from the Royal Society through the Newton Advanced Fellowship scheme (Grant no. NA140436) and IDA from CNPq (Grant: PVE-Processo: 401414/2014-0). IDA acknowledges hospitality and partial financial support by the International Institute of Physics, Federal University of Rio Grande do Norte, Natal, Brazil.
\end{acknowledgements}

\clearpage
\appendix

\section{Exact optimal free state for a four level system}\label{sec:analyticaldf}

By careful consideration of the interaction distance, we may obtain a full analytical solution for the optimal free state entanglement levels, and for $D_{\mathcal{F}}$ itself, for a four level system, $\rho^\mathrm{int}$.
A system of $N$ single-particle entanglement levels, $\lbrace \epsilon_j\rbrace$, has a $2^N\times2^N$-dimensional entanglement Hamiltonian, $\hat{H}^f_E$, with $2^N$ levels in the many-body entanglement spectrum, $\lbrace E_j\rbrace$.
Therefore, a free spectrum with four many-body levels has two single-particle levels, $\epsilon_1$ and $\epsilon_2$, that build the full spectrum.
It is convenient to work with probability densities, $\rho^\text{opt} = e^{-\hat{H}^f_E}$, allowing the single-particle energies to be reparametrised as: $b_1$ and $b_2$.
The free many-body spectrum can then be built in the following way:
\begin{align*}
\rho^\text{opt}_1 &= \left(\frac{1}{2}+b_1\right)\left(\frac{1}{2}+b_2\right),  & \rho^\text{opt}_2 &= \left(\frac{1}{2}-b_1\right)\left(\frac{1}{2}+b_2\right), \nonumber\\
\rho^\text{opt}_3 &= \left(\frac{1}{2}+b_1\right)\left(\frac{1}{2}-b_2\right),  & \rho^\text{opt}_4 &= \left(\frac{1}{2}-b_1\right)\left(\frac{1}{2}-b_2\right). \\
\end{align*}
To ensure the spectrum is normalised these levels are subject to $\sum_j \rho^\text{opt}_j=1$, along with $0\leq b_1\leq b_2\leq \frac{1}{2}$ which fixes the ordering to that of interacting spectrum: $0\leq \rho^\text{opt}_4\leq \rho^\text{opt}_3 \leq \rho^\text{opt}_2 \leq \rho^\text{opt}_1 \leq 1$.

As a first attempt to minimise the interaction distance one may directly differentiate Eq.~\eqref{eq:df} of the main text, having substituted in the definitions above,    to find the stationary points for some choice of parameters $b_1$ and $b_2$.
These parameters contain all of the information required to build the free many body entanglement spectrum, so it is our goal to find the set that minimises $D_\mathcal{F}$.
In doing so, we find that the derivatives are not defined in the regions $\rho^\mathrm{opt}_j=\rho^\mathrm{int}_j$ for any $j$.
We also find that second derivatives are always zero when $\rho^\mathrm{opt}_j\neq\rho^\mathrm{int}_j$, thus defining a saddle point and not a minimum.
The minimum trace distance must therefore live on one of the boundary curves $\rho^\mathrm{opt}_j=\rho^\mathrm{int}_j$ or an intersection of two or more curves.
Of course, it is the low-level entanglement spectrum that provides important information about the system.
Therefore, if it is possible to match the low levels then the optimal free state will more faithfully represent the interacting system.
In some cases, however, the intersection between the low level curves does not fall within the normalised and ordered region.
In that case the most faithful representation lies on the curve $b_1=b_2$.

\begin{figure}[t]
\centering
\includegraphics[width=0.4\paperwidth]{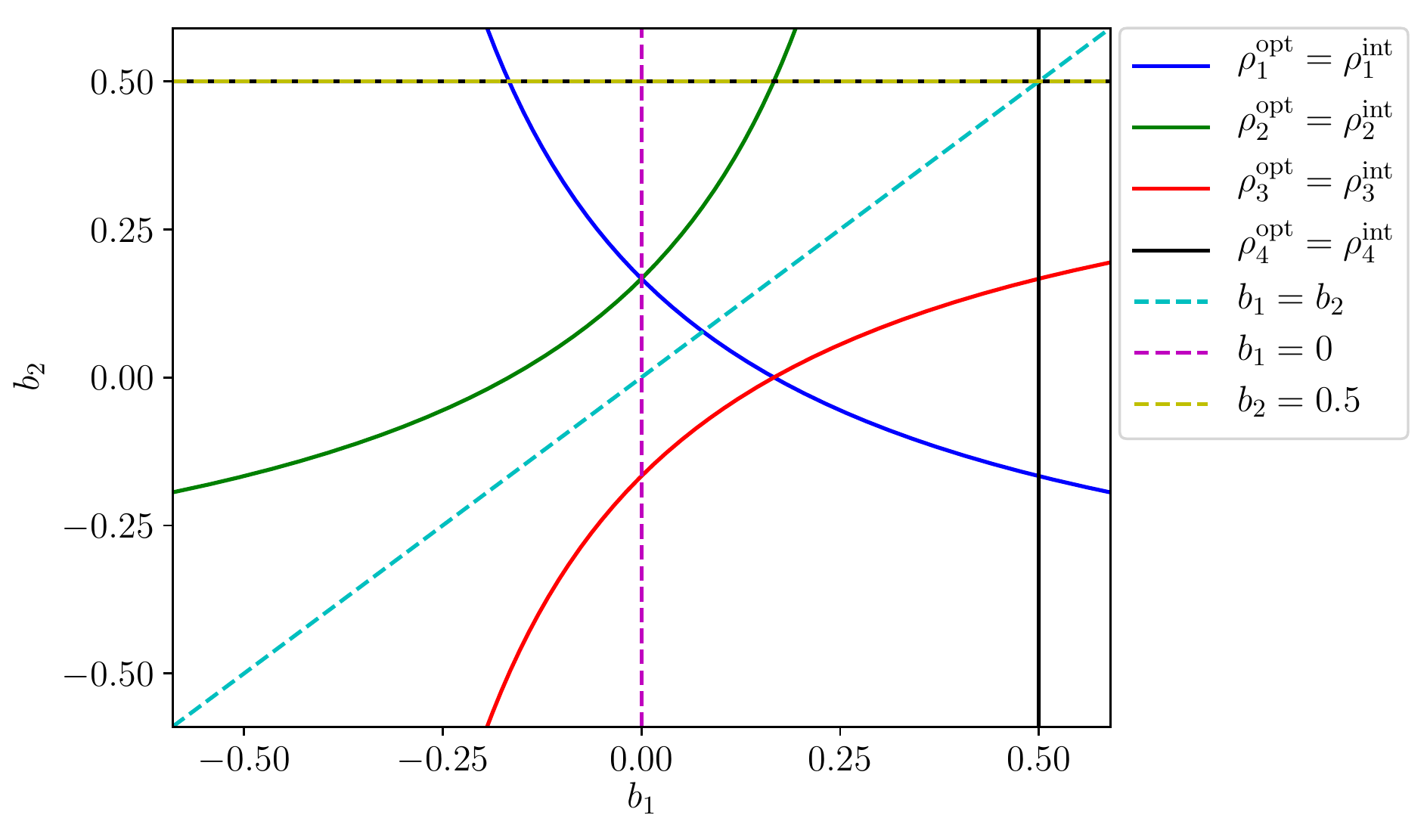}
\caption{
Free parameter values, $b_1,b_2$, that produce boundary curves $\rho_j^\mathrm{opt}=\rho_j^\mathrm{int}$ (solid lines) for $\lbrace\rho_j^\mathrm{int}\rbrace=\lbrace \frac{1}{3},\frac{1}{3},\frac{1}{3},0\rbrace$.
The dashed lines enclose the normalised and ordered regions for the free spectra.
There are two points of intersection within the normalised and ordered region.
The intersection that matches better the low level entanglement spectrum will give the most faithful representation of the interacting system.
Thus, it is the $b_1,b_2$ pair at the intersection between $\rho^\mathrm{opt}_1=\rho^\mathrm{int}_1$ and $\rho^\mathrm{opt}_2=\rho^\mathrm{int}_2$ that give the interaction distance, $D_\mathcal{F}=\frac{1}{6}$.
}
\label{fig:boundarycurve}
\end{figure}

An exhaustive analysis yields the following set of solutions for the interaction distance, where the superscript `int' has now been dropped on all $\rho^\mathrm{int}_j$:
\begin{numcases}{D_{\mathcal{F}}  =}
2\sqrt{\rho_1} - 2\rho_1- \rho_2- \rho_3, & $\text{if  } \rho_1\geq(\rho_1+\rho_2)^2 $ \notag\\
 & \label{eq:df1}\\
\left| \frac{\rho_1\rho_4-\rho_2\rho_3}{\rho_1+\rho_2}\right|, & $\text{otherwise}.$ \label{eq:df2}
\end{numcases}
and the following set of free parameter solutions:
\begin{numcases}{(b_1,b_2) = }
\left(\sqrt{\rho_1} - \frac{1}{2},  \sqrt{\rho_1} - \frac{1}{2} \right),& \hspace{-3mm}$\text{if  } \rho_1\geq(\rho_1+\rho_2)^2$ \notag\\
 & \label{eq:b1}\\
\left(\frac{\rho_1-\rho_2}{2(\rho_1+\rho_2)},  \rho_1+\rho_2-\frac{1}{2} \right),&\hspace{-3mm} $\text{otherwise}.$ \notag\\
&\label{eq:b2}
\end{numcases}
These exact solutions allow for an accurate study of the interaction distance without any error of numerical optimisation.
The solutions~\eqref{eq:df1} and~\eqref{eq:df2} correspond to the cases where it is not possible and possible to match the lowest two levels of the entanglement spectrum, respectively.

In Fig.~\ref{fig:boundarycurve} we show an example of the boundary curves for the interacting spectrum $\lbrace \rho_j\rbrace=\lbrace \frac{1}{3},\frac{1}{3},\frac{1}{3},0\rbrace$ that produces $D_\mathcal{F}=\frac{1}{6}$.
We are able to deduce this solution by first considering the condition: $\rho_1\geq(\rho_1+\rho_2)^2$.
For our set $\lbrace\rho_j\rbrace$, this inequality is not satisfied so $b_1\neq b_2$ and the minimum trace distance must therefore live at an intersection between the $\rho_1^\mathrm{opt}=\rho_1$ and $\rho_2^\mathrm{opt}=\rho_2$ curves.
The pair $b_1,b_2$ at this intersection result in the interaction distance.

\section{Closed form solution of $D_\mathcal{F}$ for Hubbard Dimer}

In agreement with the result in Ref. \cite{Carrascal}, for the Hubbard dimer with Hamiltonian~\eqref{eq:hamiltonian} of the main text (with $H\to H-(\nu_1+\nu_2)$ and $J\neq0$), we find the ground state energy in the half-filled, $S_z=0$ sector to be
\begin{align}
E = -\frac{2}{3}A\cos\left(\theta\right) + \frac{2U}{3}
\end{align}
with $A= \left[U^2 + 3\Delta\nu^2 + 12 J^2\right]^\frac{1}{2}$ and $\cos\left(3\theta\right) = \frac{U(36J^2 - 18\Delta\nu^2 +2U^2)}{2(12J^2 + 3\Delta\nu^2 +U^2)^\frac{3}{2}}$. The corresponding state is
\begin{align}\label{eq:exactgs}
\ket{\psi_0}&= \frac{1}{N^\frac{1}{2}}\left[ 2J \ket{\uparrow\downarrow,0} + a\ket{\uparrow,\downarrow}- a\ket{\downarrow,\uparrow} + 2J\frac{a}{b}\ket{0,\uparrow\downarrow}\right]^\frac{1}{2}
\end{align}
with $a = (U+\Delta\nu - E)$, $b=(U - \Delta\nu - E)$ and $N = \left[4J^2 +2a^2 + 4J^2\frac{a^2}{b^2}\right]$. 

From the ground state, Eq.~\eqref{eq:exactgs}, it is possible to extract the entanglement spectrum of the interacting model. This can be used together with Eq.'s~\eqref{eq:df1}-\eqref{eq:b2} to produce an exact optimal free entanglement spectrum and the correpsonding interaction distance.
In the strongly correlated regime, when $U\gg J,\Delta\nu$, the interaction distance is
\begin{align}
D_\mathcal{F} =\frac{2J^2}{N} \left| \frac{b^2 - a^2 }{b^2}\right|.
\end{align}

\end{document}